\documentclass[a4paper,11pt]{article}
\pdfoutput=1 % if your are submitting a pdflatex (i.e. if you have
\usepackage{jheppub} % for details on the use of the package, please
\usepackage[T1]{fontenc} % if needed

\usepackage{CJK}
\usepackage{xcolor}
\usepackage{amsfonts}
\usepackage{epstopdf}
\usepackage{wrapfig} % to wrap figure with texts

\def\beq{\begin{equation}}
\def\eeq{\end{equation}}

\def\bc{\begin{center}}

\def\ec{\end{center}}
\def\be{\begin{eqnarray}}
\def\ee{\end{eqnarray}}

\title{Holographic Josephson Junction from Massive Gravity}

\author[a,e]{Ya-Peng Hu,}
\author[b]{Huai-Fan Li,}
\author[c]{Hua-Bi Zeng,}
\author[d]{and Hai-Qing Zhang}
\affiliation[a]{College of Science, Nanjing University of Aeronautics and Astronautics,\\ Nanjing 210016, China;}
\affiliation[b]{Institute of Theoretical Physics, Department of Physics, Shanxi Datong University, \\Datong 037009, China;}
\affiliation[c]{Department of Physics, National Central University, Chungli 320, Taiwan;}
\affiliation[d]{Institute for Theoretical Physics, Utrecht University, \\Leuvenlaan 4, 3584 CE Utrecht, The Netherlands}
\affiliation[e]{State Key Laboratory of Theoretical Physics, Institute of Theoretical Physics, Chinese Academy of Sciences, Beijing, 100190, China}
\emailAdd{huyp@nuaa.edu.cn}
\emailAdd{huaifan.li@stu.xjtu.edu.cn}
\emailAdd{zenghbi@gmail.com}
\emailAdd{H.Q.Zhang@uu.nl}

\abstract{We study the holographic superconductor-normal metal-superconductor (SNS) Josephon junction in the massive gravity. In the homogeneous case of the chemical potential, we find that the graviton mass will make the normal metal-superconductor phase transition harder to take place. In the holographic model of Josephson junction, it is found that the maximal tunneling current will decrease according to the graviton mass. Besides, the coherence length of the junction decreases as well with respect to the graviton mass. If one interprets the graviton mass as the effect of momentum dissipation in the boundary field theory, it indicates that the stronger the momentum dissipation is, the smaller the coherence length is.  }

\begin{document}
\maketitle
\flushbottom
\begin{CJK*}{GBK}{song}

\section{Introduction}
\label{sect:intro}
AdS/CFT correspondence \cite{Maldacena:1997re, Gubser:1998bc, Witten:1998qj} is a powerful tool to solve the strongly coupled physics from the one dimensional higher weakly coupled gravity in some limit, which is also dubbed `holography'. In recent years, the applied holography has attained great success in the area of condensed matter theory (CMT), hydrodynamics and etc. \cite{Hartnoll:2008vx, Lee:2008xf, Kovtun:2004de}. The holographic Josephson junction was first studied in \cite{Horowitz:2011dz}, later on was largely extended to various models \cite{Wang:2011rva}. Other models of holographic Josephson junction were proposed from designer multi-gravity in \cite{elias} and from D-branes in \cite{hoyos}.

The gravity backgrounds in the above Josephson junctions are mostly the Einstein's general relativity (GR). Due to the requirements of diffeomorphism invariance in GR, the corresponding graviton should be a massless spin-2 boson, see for example~\cite{tony}. Therefore, it is a direct and interesting task in history to find whether there are other gravitational theories in which the graviton is massive~\cite{Fierz:1939ix}. However, the generalization is not so easy since usually the massive gravity has the instability problem of the Boulware-Deser ghost~\cite{Boulware:1973my}. Recently, a nonlinear massive gravity theory has been proposed (the so-called dRGT theory by de Rham, Gabadadze, and Tolley)\cite{deRham:2010ik, deRham:2010kj, Hinterbichler:2011tt}, and later it is found to be ghost-free~\cite{Hassan:2011hr,Hassan:2011tf}. For more details about the aspects of massive gravity, one can refer to the reviews~\cite{Hinterbichler:2011tt,deRham:2014zqa}. There have been many investigations on this type of massive gravity, for instance, the black hole solutions and their thermodynamics were studied in \cite{Vegh:2013sk,Adams:2014vza,Cai:2014znn,Hu:2015xva,Xu:2015rfa,Hendi:2015hoa}. The counterterm of this massive gravity has been obtained in \cite{Cao:2015cza},  as well as that it has been proved to be ghost-free for this massive gravity with a special degenerate reference metric in~\cite{Zhang:2015nwy}. Moreover, due to the breaking of diffeomorphism invariance in dRGT massive gravity, the stress energy tensor of matter is not conserved anymore. The non-conservation of stress energy tensor is dual to the dissipation of momentum in the boundary field theory according to the AdS/CFT, such that a finite DC conductivity was obtained in~\cite{Vegh:2013sk,Davison:2013jba,Blake:2013bqa,Blake:2013owa}. Some other holographic results related to the effects of graviton mass in massive gravity have also been investigated in~\cite{Davison:2013txa,Adams:2014vza,Amoretti:2014zha,Zeng:2014uoa,Baggioli:2015zoa}.

In this paper, we are going to investigate the effect of the graviton mass (or breaking of translational symmetry in the boundary field theory due to AdS/CFT \cite{Blake:2013bqa}) on the holographic Josephson junction. In particular, we study the Superconductor-Normal metal-Superconductor (SNS) Josephson juction in the dRGT massive gravity. First, for a homogeneous superconductor, we find that the graviton mass will reduce the critical temperature from a normal metal to a superconductor, which indicates it is harder to have a  phase transition from normal metal to a superconductor when graviton mass is larger.  Meanwhile, from the AdS/CFT aspect it also means in the boundary field theory, the larger the momentum dissipation is, the harder the phase transition is. This is reminiscent of the phase diagram of cuprates, where greater doping makes the phase transition from normal metal to superconductor more difficult to take place. Although doping is not the same as the momentum dissipation, we can still argue that they have similar effects to the phase transitions from the point view of holography.
For the holographic SNS Josephson junction, we find the usual sinusoidal relation between the tunneling current the phase difference across the junction. One can obtain the maximal current by fitting the sinusoidal relation. It is found that the maximal current decreases exponentially with width of the junction, from which we can get the coherence length of the normal metal in-between the junction \cite{tinkham}. Moreover, we find that the coherence length decreases with respect to the graviton mass, so does the maximal current. Physically, it indicates that the momentum dissipation (or breaking of translational symmetry) will reduce the coherence length as well as the maximal current.

The paper is organized as follows. In Sec.\ref{sect:MG} we will briefly introduce the dRGT massive gravity; We build up the holographic model of Josephson junction in Sec.\ref{sect:setup}; Numerical results are shown in Sec.\ref{sect:result} and conclusions are drawn in Sec.\ref{sect:con}; In particular, the dependence of the constraint equation with the gauge field equations are generically proven in Appendix \ref{sect:appendix}.

\section{A Brief Review: dRGT Massive Gravity and Its General Black Hole Solutions}
\label{sect:MG}
In this paper, we will focus on the ghost-free dRGT massive gravity, whose action in an $(n+2)$-dimensional spacetime is usually read as~\cite{Vegh:2013sk,Cai:2014znn}
\begin{equation}
\label{actionmassive}
S =\frac{1}{16\pi G}\int d^{n+2}x \sqrt{-g} \left[ R +\frac{n(n+1)}{L^2} +m^2 \sum^4_i c_i {\cal U}_i (g,\mathfrak{f})\right],
\end{equation}
where $\mathfrak{f}$ is a fixed symmetric tensor usually called the reference metric, $L$ is the radius of AdS$_{n+2}$ spacetime;
$c_i$ are constants,  $m$ stands for the graviton mass,\footnote{Precisely, $m$ is the graviton mass near the UV boundary. From \cite{Blake:2013bqa} we know that the effective graviton mass depends on the radial direction. By taking the radial direction to the UV boundary, we can see that $m$ here is proportional to the graviton mass near the UV boundary up to a constant. For simplicity, we call $m$ here as graviton mass, the exact meaning of which is clear from the above explanation.} and ${\cal U}_i$ are symmetric polynomials of the eigenvalues of the $(n+2)\times (n+2)$ matrix ${\cal K}^{\mu}_{\ \nu} \equiv \sqrt {g^{\mu\alpha}\mathfrak{f}_{\alpha\nu}}$:
\begin{eqnarray}
\label{eq2}
&& {\cal U}_1= [{\cal K}], \nonumber \\
&& {\cal U}_2=  [{\cal K}]^2 -[{\cal K}^2], \nonumber \\
&& {\cal U}_3= [{\cal K}]^3 - 3[{\cal K}][{\cal K}^2]+ 2[{\cal K}^3], \nonumber \\
&& {\cal U}_4= [{\cal K}]^4- 6[{\cal K}^2][{\cal K}]^2 + 8[{\cal K}^3][{\cal K}]+3[{\cal K}^2]^2 -6[{\cal K}^4].
\end{eqnarray}
The square root in ${\cal K}$ means $(\sqrt{A})^{\mu}_{\ \nu}(\sqrt{A})^{\nu}_{\ \lambda}=A^{\mu}_{\ \lambda}$ and $[{\cal K}]=K^{\mu}_{\ \mu}=\sqrt {g^{\mu\alpha}\mathfrak{f}_{\alpha\mu}}$. After making variation of action with respect to the metric, the equations of motion (EoM) turns out to be
\begin{eqnarray}
R_{\mu\nu}-\frac{1}{2}Rg_{\mu\nu}-\frac{n(n+1)}{2L^2} g_{\mu\nu}+m^2 \chi_{\mu\nu}&=&8\pi G T_{\mu \nu },~~
\end{eqnarray}
where
\begin{eqnarray}
&& \chi_{\mu\nu}=-\frac{c_1}{2}({\cal U}_1g_{\mu\nu}-{\cal K}_{\mu\nu})-\frac{c_2}{2}({\cal U}_2g_{\mu\nu}-2{\cal U}_1{\cal K}_{\mu\nu}+2{\cal K}^2_{\mu\nu})
-\frac{c_3}{2}({\cal U}_3g_{\mu\nu}-3{\cal U}_2{\cal K}_{\mu\nu}\nonumber \\
&&~~~~~~~~~ +6{\cal U}_1{\cal K}^2_{\mu\nu}-6{\cal K}^3_{\mu\nu})
-\frac{c_4}{2}({\cal U}_4g_{\mu\nu}-4{\cal U}_3{\cal K}_{\mu\nu}+12{\cal U}_2{\cal K}^2_{\mu\nu}-24{\cal U}_1{\cal K}^3_{\mu\nu}+24{\cal K}^4_{\mu\nu}).~~
\end{eqnarray}
Since the background we are going to use is $(3+1)$ dimension, thus a general black hole solution can be~\cite{Cai:2014znn}
\begin{eqnarray}\label{metric}
ds^2&=&-r^{2}f(r)dt^2+\frac{dr^2}{r^2 f(r)}+r^2h_{ij}dx^idx^j,\\
\label{fr} f(r)&=&\frac{k}{r^2}+\frac{1}{L^2}-\frac{m_0}{r^3}+\frac{q^2}{4r^4}+\frac{c_1m^2}{2r}+\frac{c_2m^2}{r^2},
\end{eqnarray}
where $h_{ij}dx^idx^j$ is the line element for the 2-dimensional spherical, flat or hyperbolic space with $k=1,~0$ or $-1$ respectively. $m_0$ is related to the mass of the black hole while $q$ is the charge of it. The reference metric now can have a special choice
\begin{equation}
\mathfrak{f}_{\mu\nu}=\text{diag}~\{0,0,h_{ij}\}.
\end{equation}
The Hawking temperature of this black hole solution can be easily found
\be
T_{BH}=\frac{\left(r^2f(r)\right)'}{4\pi}\bigg|_{r=r_+}=\frac{1}{4\pi r_+}\left(k+\frac{3r_+^2}{L^2}-\frac{q^2}{4r_+^2}+c_1m^2r_++c_2m^2\right).
\ee
in which $r_+$ is the horizon of the black hole.

\section{Holographic Setup}
\label{sect:setup}
For simplicity, we will consider the black hole solution in (\ref{metric}) with $k=0$ and $q=0$, therefore, $h_{ij}=\text{diag}(h_{xx},h_{yy})=\text{diag}(1,1)$.
In the probe limit, we adopt the Maxwell and complex scalar field action as
\begin{equation}\label{action2}
S=\int d^{4} x\sqrt{-g}\left(-\frac{1}{4}F_{\mu\nu}F^{\mu\nu}-|\nabla\psi-iA\psi|^2-m_\psi^2|\psi|^2\right),
\end{equation}
 in which $A_\mu$ is the $U(1)$ gauge field while $F_{\mu\nu}$ is the  corresponding field strength with $F_{\mu\nu}=\partial_\mu A_\nu-\partial_\nu A_\mu$; $m_\psi$ is the mass of the complex scalar field $\psi$. The EoMs can be obtained readily from the above action as
\be \label{eomA}
0&=&(\nabla_\mu-iA_\mu)(\nabla^\mu-iA^\mu)\psi-m_\psi^2\psi, \\
\label{eompsi}\nabla_\nu F^{\nu\mu}&=&i(\psi^*(\nabla^\mu-iA^\mu)\psi-\psi(\nabla^\mu+iA^\mu)\psi^*).
\ee
In order to work with the gauge-invariant fields, a natural ansatz for the fields are
\begin{equation}\label{ansatz}
\psi=|\psi|e^{i\varphi},\quad A_\mu=(A_t, A_r, A_x, 0),
\end{equation}
where $|\psi|,\varphi,A_t,A_r,A_{x}$ are all real functions of $r$ and $x$. The corresponding gauge-invariant quantity can be defined as $M_{\mu}\equiv A_{\mu}-\partial_{\mu}\varphi$. Therefore, from the formula \eqref{metric}, \eqref{eomA}, \eqref{eompsi} and \eqref{ansatz}, we obtain the following coupled partial differential equations (PDEs)~:
\begin{subequations}
\be
\partial_r^2|\psi|+\frac{1}{r^4 f}\partial_x^2|\psi|+\left(\frac{4}{r}+\frac{f'}{f}\right)\partial_r|\psi|+\frac{1}{r^2 f}\left(\frac{M_t^2}{r^{2}f}-r^2 f M_r^2-\frac{M_x^2}{r^2}-L^2 m_\psi^2\right)|\psi|&=&0,  \quad\quad\quad\label{eom1}\\
\partial_r M_r+\frac{1}{r^4 f}\partial_x M_x+\frac{2}{|\psi|}\left(M_r\partial_r|\psi|+\frac{M_x}{r^4 f}\partial_x|\psi|\right)+\left(\frac{4}{r}+\frac{f'}{f}\right)M_r&=&0, \label{eom2}\\
\partial_r^2 M_t+\frac{1}{r^4 f}\partial_x^2 M_t+\frac{2}{r}\partial_r M_t-\frac{2L^2|\psi|^2}{r^2 f}M_t&=&0, \label{eom3}\\
\partial_x^2 M_r-\partial_x\partial_r M_x-2L^2 r^2 |\psi|^2 M_r&=&0, \label{eom4}\\
\partial_r^2 M_x-\partial_x\partial_r M_r+\left(\frac{f'}{f}+\frac{2}{r}\right)\left(\partial_r M_x-\partial_x M_r\right)-\frac{2|\psi|^2}{L^2 r^2 f}M_x&=&0. \label{eom5}
\ee\label{pdes}
\end{subequations}
where $'$ denotes the derivative with respect to $r$. One can find that only gauge-invariant quantities are left in the above PDEs, and the phase $\varphi$ has been absorbed into the gauge-invariant quantity $M_\mu$. Moreover, there are only four independent EoMs in the above five EoMs, since the second equation \eqref{eom2} is just a constraint equation which results from the algebraic combinations of \eqref{eom4} and \eqref{eom5} as \footnote{In the Appendix \ref{sect:appendix}, we show how the constrained equation can be obtained from the Maxwell equation in a generic case. }
\be
\text{Eq.}\eqref{eom2}=-\frac{1}{2r^2|\psi|^2}\left(\partial_r[\text{Eq.}\eqref{eom4}]
+\left(\frac{f'}{f}+\frac2r\right)\times\text{Eq.}\eqref{eom4}
+\partial_r[\text{Eq.}\eqref{eom5}]\right).
\ee
Therefore, we will correctly work with four independent EoMs for four fields, {\it i.e.}, $|\psi|, M_t, M_r$ and $M_x$.

In order to solve the above coupled EoMs, we first need to impose boundary conditions for these fields. At the horizon, the field $M_t$ should vanish since $g^{tt}$ is divergent at horizon,  while other fields are finite at horizon.

Near the infinite boundary $r\to\infty$, the fields $|\psi|, M_r$ and $M_x$ can be expanded as,
\begin{equation} \label{uv}
\begin{split}
|\psi|=&\frac{\psi^{(1)}(x)}{r^{(3-\sqrt{9+4m_\psi^2})/2}}+\frac{\psi^{(2)}(x)}{r^{(3+\sqrt{9+4m_\psi^2})/2}}
+\mathcal{O}(\frac{1}{r^{(5+\sqrt{9+4m_\psi^2})/2}}), \\
M_t=&\mu(x)-\frac{\rho(x)}{r}+\mathcal{O}(\frac{1}{r^{2}}), \\
M_r=&\frac{M_r^{(1)}(x)}{r^{2}}+\mathcal{O}(\frac{1}{r^{3}}), \\
M_x=&\nu(x)+\frac{J(x)}{r}+\mathcal{O}(\frac{1}{r^{2}}).
\end{split}
\end{equation}
According to the AdS/CFT correspondence, the scalar field $|\psi|$ has conformal dimension $\Delta_\pm=(3\pm\sqrt{9+4m_\psi^2})/2$. In the following, we will impose $\psi^{(1)}\equiv0$, which indicates there is no source term of the scalar operator on the boundary.  According to the AdS/CFT dictionary, the coefficients $\psi^{(2)}$, $\mu$, $\rho$, $\nu$ and $J$ correspond to the condensate of the dual scalar operator $\langle \mathcal{O}\rangle$, chemical potential, charge density, superfluid velocity and current in the boundary field theory, respectively.%~\footnote{We also notice that there is a relation $\partial_x^2 M_r^{(1)}(x)+\partial_xJ(x)=0$, which can be used to set $J$=const  by imposing $\partial_x M_r^{(1)}=0$, in the numerical calculations in the next section. }
 The gauge-invariant phase difference $\gamma=\Delta \varphi-\int A_x$ across the weak link can be defined as \cite{Horowitz:2011dz}
\be\label{gamma}
\gamma=-\int^{+\infty}_{-\infty}dx[\nu(x)-\nu(\pm\infty)].
\ee
where $\nu(\pm\infty)$ is the superfluid velocity at the spacial infinity $x=\pm\infty$.

In order to model a SNS Josephson junction, we choose the chemical potential $\mu(x)$ as
\begin{equation}\label{mu}
\mu(x)=\mu_\infty\left\{1-\frac{1-\epsilon}{2\tanh(\frac{\ell}{2\sigma})}\left[\tanh(\frac{x+\frac{\ell}{2}}{\sigma})-\tanh(\frac{x-\frac{\ell}{2}}{\sigma})\right]\right\},
\end{equation}
where $\mu_\infty=\mu(+\infty)=\mu(-\infty)$ is the chemical potential at $x=\pm\infty$, while $\ell$, $\sigma$ and $\epsilon$ are the width, steepness and depth of the junction, respectively. Following ref.~\cite{Horowitz:2011dz}, we can define the critical temperature of the Josephson junction $T_c$  as the critical temperature of a homogenous superconductor, {\it i.e.}, with a flat chemical potential. Therefore,  $T_c$ is proportional to $\mu_\infty=\mu(\pm\infty)$:
\begin{equation}\label{tem1}
T_c=\frac{T_{BH}}{\mu_c}\mu(\infty),
\end{equation}
where $\mu_c$ is the critical chemical potential of a homogenous superconductor without any current at temperature  $T_{BH}$. Inside the junction, $x\in(-\frac{\ell}{2},\frac{\ell}{2})$, the effective critical temperature can be defined as
\begin{equation}\label{tem2}
T_0=\frac{T_{BH}}{\mu_c}\mu(0),
\end{equation}
Therefore, if one can set the profile of the chemical potential as $\mu(0)<\mu_c<\mu(\infty)$, from the relations \eqref{tem1} and \eqref{tem2} we know that $T_0<T_{BH}<T_c$. Hence, the in-between junction is in the normal metallic phase, while the region outside the junction is in superconducting phase. In the following section, we will work in this way to model the SNS Josephson junction.

\section{Numerical Results}
\label{sect:result}

There is a scaling symmetry in the PDEs~\eqref{pdes}:
\begin{equation}\label{scal}
 t\rightarrow\lambda t,\quad x\rightarrow\lambda x, \quad y\rightarrow\lambda y, \quad r\rightarrow\frac{1}{\lambda}r,\quad M_t\rightarrow\frac{1}{\lambda}M_t,\quad M_x\rightarrow \frac{1}{\lambda} M_x,\quad M_r\rightarrow \lambda M_r,
\end{equation}
where $\lambda$ is an arbitrary constant. We can adopt the above scaling symmetry \eqref{scal} to set the horizon $r_+=1$ in the numerics. For convenience, we use the transformed coordinates in the following way $u=1/r$ and $y=\tanh(\frac{x}{4\sigma})$, as well as
\be\label{uv2}
\begin{split}
|\psi|\rightarrow \frac{|\psi|}{r^{\left(3-\sqrt{9+4m_\psi^2}\right)/2}},\quad\quad
M_r\rightarrow\frac{M_r}{r^{2}}.
\end{split}
\ee
Without loss of generality, we set the AdS radius $L=1$. We choose $m_\psi^2=-2$ in the numerics and the range of graviton mass are $0\leq m\leq 1.2$,  since in the following we find that at $m\sim1.2$ the maximal current is extremely close to zero as the width of the junction is large.   For the convenience of numerics, we set $c_1=1, c_2=-1/2$ in \eqref{fr} in order to fix the horizon at $r_+=1$ by varying the mass of graviton $m$.\footnote{According to \cite{Cai:2014znn}, the background in $(3+1)$ dimensions with $k=0$ is thermodynamically stable for $c_2\leq0$.} We solve the EoMs~\eqref{eom1}-\eqref{eom5} numerically by means of Chebyshev spectral methods \cite{tinkham}.

\subsection{Critical Temperature}
%%%%%%%%%%
\begin{figure}[htbp]
\centering
\includegraphics[trim=0cm 1.cm 0cm 0.5cm, clip=true,scale=0.245]{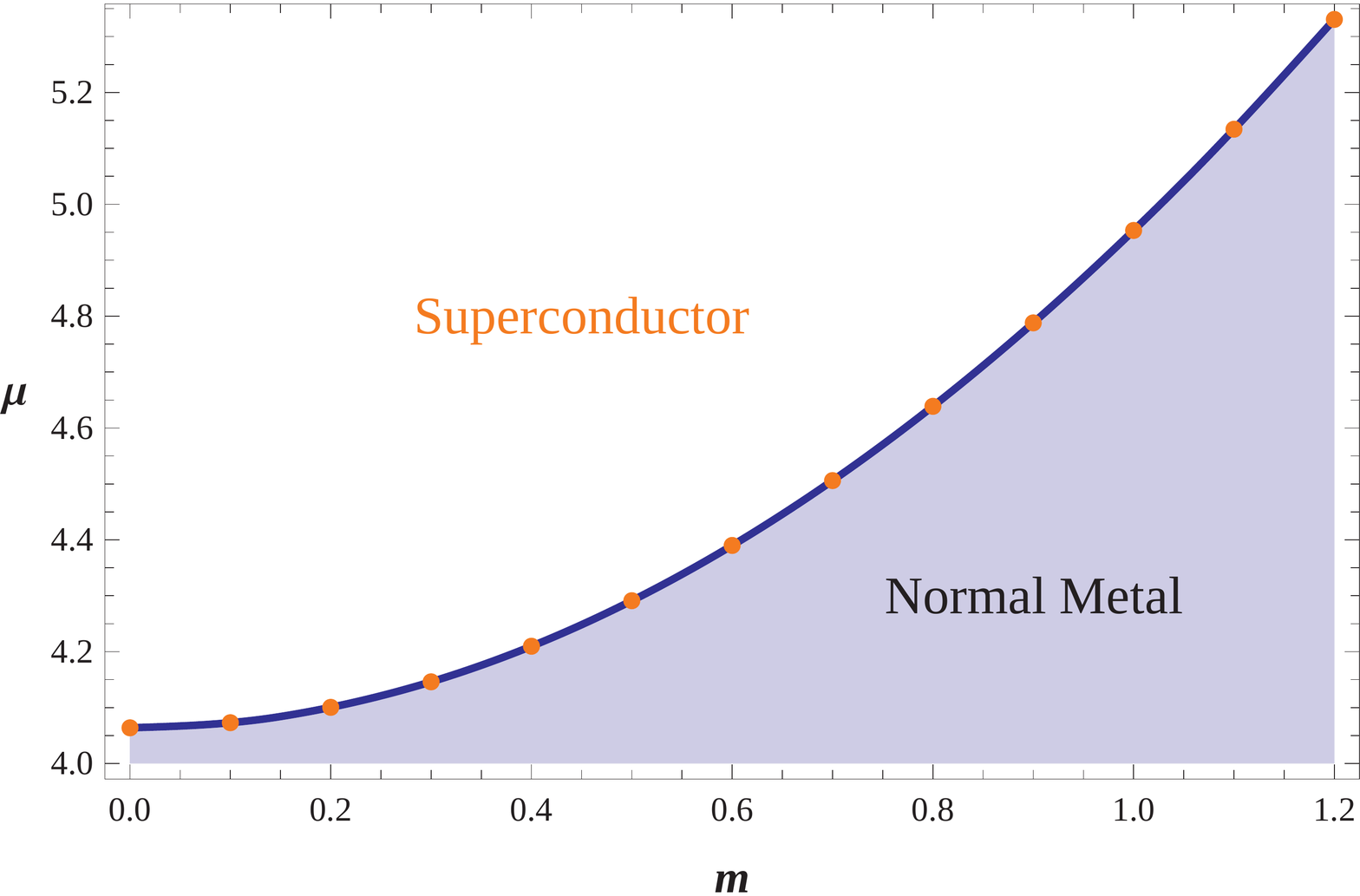}
\includegraphics[trim=0cm 1.5cm 0cm 1.3cm, clip=true,scale=0.26]{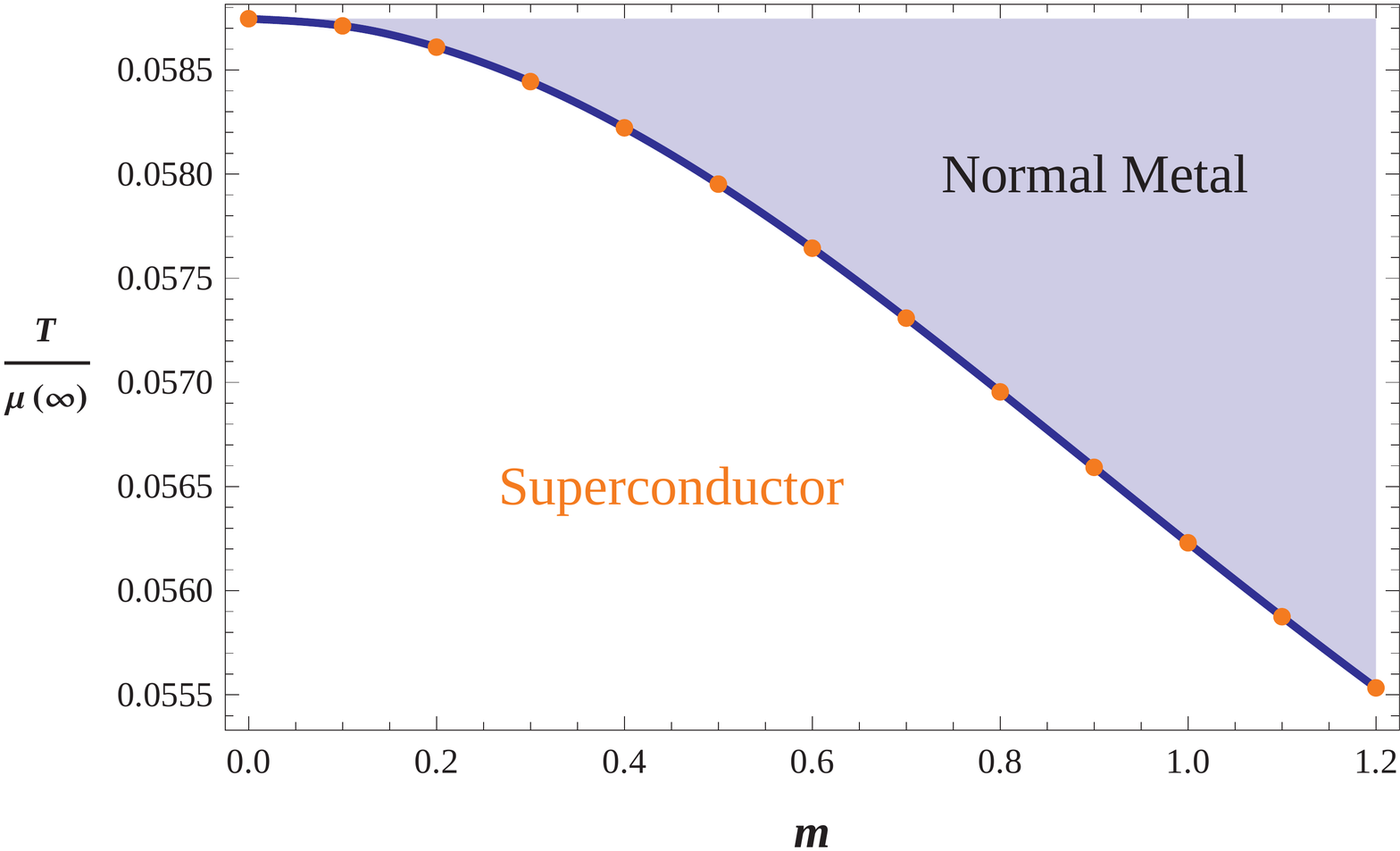}
\caption{\label{tcmass} Phase diagrams of superconductor and normal metal. (Left) Chemical potential versus graviton mass $m$; (Right) Temperature versus graviton mass $m$. Orange dots are from numerical data. }
\end{figure}
In this subsection, we will study the phase diagram of the boundary theory with homogeneous chemical potential.  The  critical chemical potentials $\mu_c$ from normal metal states to superconductor states are from $\mu_c\approx4.0638$ at $m=0$ to $\mu_c\approx 5.3306$ at $m=1.2$, which are shown in the left panel of Fig.\ref{tcmass}. The phase diagram corresponds to the critical temperatures are plotted in the right panel of Fig.\ref{tcmass}. The dark region is the normal metal phase while the white region is the superconductor phase. On one hand, for a fixed graviton mass $m$, lowering temperature (increasing chemical potential) will change the normal metal state to a superconductor state; On the other hand, for a fixed temperature (chemical potential), increasing graviton mass $m$  will destroy the superconductor phase into a normal metal phase.
% \begin{wrapfigure}{r}{0.4\textwidth}
%  \begin{center}
%    \includegraphics[trim=3cm 1.5cm 3cm .3cm, clip=true,scale=0.25]{phase.pdf}
%  \end{center}
%  \caption{\label{cuprates}Phase diagram of cuprates excerpted from \cite{varma}. }
%\end{wrapfigure}
The phase diagram in Fig.\ref{tcmass} is reminiscent of the famous phase diagram in the cuprates with doping, such as the Fig.1 in \cite{varma}. Between the phases of superconductivity and Fermi liquid, greater doping will change a superconductor to a Fermi liquid or normal metal at a fixed temperature. Therefore, from this point of view there is a subtle relationship between the graviton mass and the doping. We cannot make any definite conclusion of this relationship currently, however, at least they more or less have a similar effect to the phase transition from superconductivity to normal metal. A more complicated study of this phase transition has been brought out in \cite{Baggioli:2015zoa}, where they have adopted a different action and metric from ours.

In order to model a SNS Josephson junction, from the above discussion we need to set $\mu(0)<\mu_c<\mu(\infty)$ for various $m$. After some trials, we find that a unified chemical potential $\mu(x)$  with the parameters $\mu_\infty=6, \sigma=0.7$ and $\epsilon=0.6$ will satisfy the requirements of SNS junction. We also choose $2\leq \ell\leq5$ in order to study the coherence length $\xi$ of the junction.

\subsection{Tunneling Current}
\begin{figure}[htbp]
\centering
\includegraphics[trim=0cm 2cm 0cm 2cm, clip=true,scale=0.3]{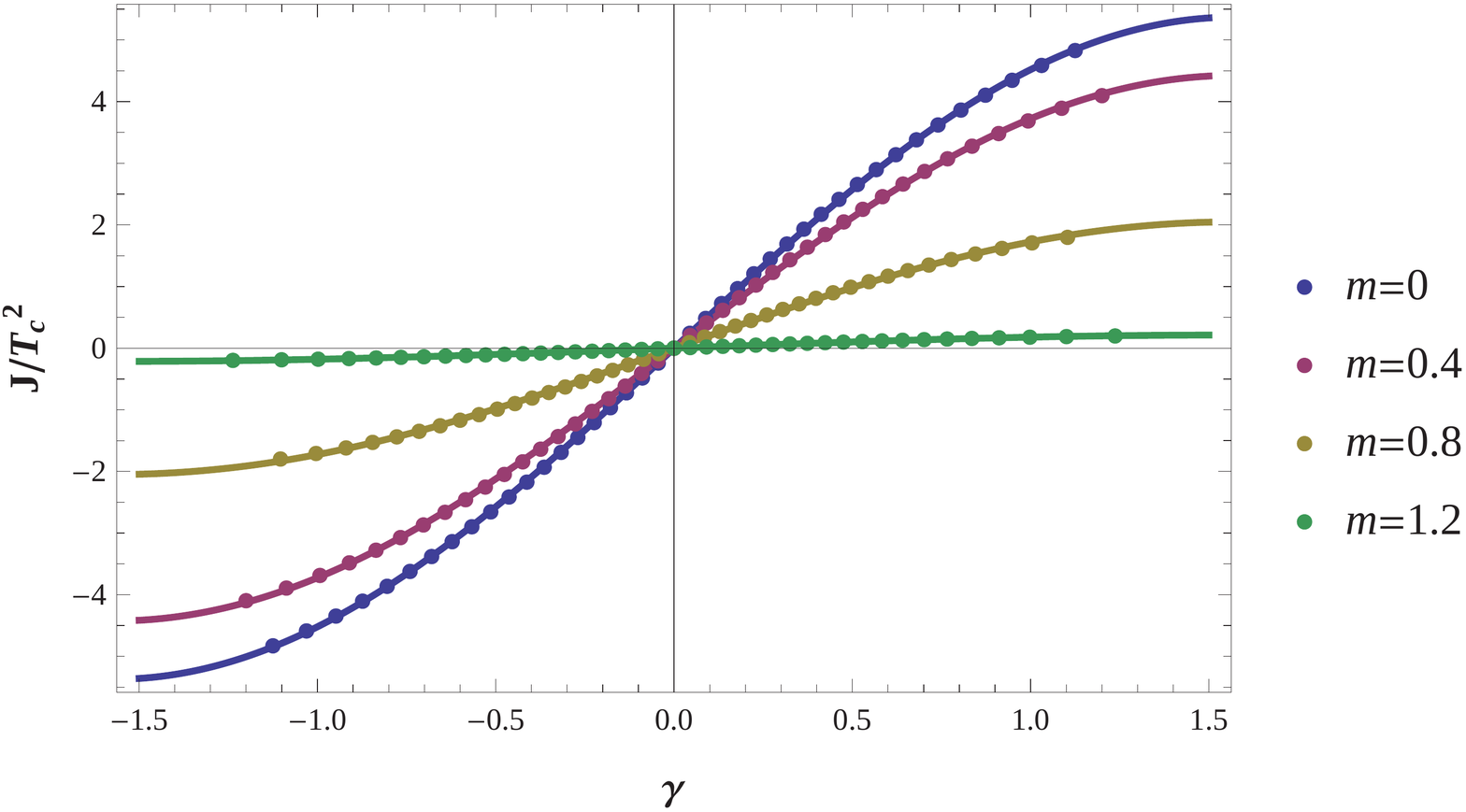}
\caption{\label{Jgamma} The sinusoidal relation between the tunneling current $J/T_c^2$ and the phase difference $\gamma$ for various $m$. The width of junction now is $\ell=2$. Dots are from the numerical data while the solid lines are the fittings of the data. }
\end{figure}
Now we are going to set the extra boundary conditions for the Josephson junctions. Close to the spatial boundary $x=\pm\infty$, we demand that all the fields are homogeneous, {\it i.e.} $\partial_x (\text{fields})=0$. There is another symmetry of the fields when we flip the sign of $x\to -x$,
\be
|\psi|\to|\psi|,\quad M_t\to M_t,\quad M_r\to-M_r,\quad M_x\to M_x.
\ee
Therefore, $M_r$ is an odd function of $x$ while others are even. Thus it is natural to set $M_r(x=0)=0$, while other fields have vanishing first order derivative with respect to $x$ at $x=0$. In the numerics, we set $J$ as a constant and make it as an input parameter. Hence, the velocity $\nu$ and phase difference $\gamma$ can be obtained numerically.  Moreover, we find that it is convenient to work with dimensionless quantities, for instance $J/T_c^2$.

\begin{figure}[htbp]
\centering
\includegraphics[trim=0cm 1.5cm 0cm 1.3cm, clip=true,scale=0.27]{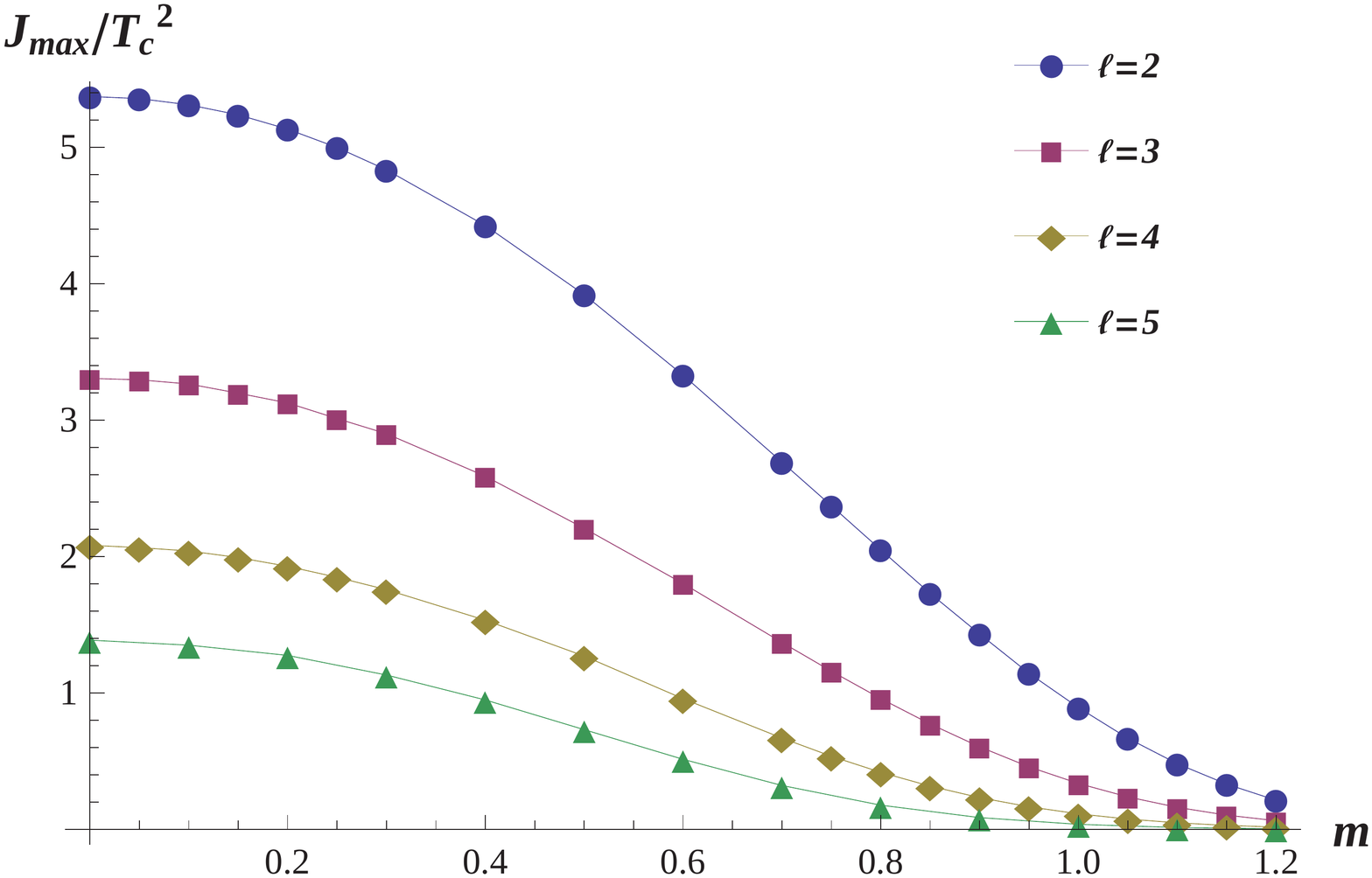}
\caption{\label{Jmax} Maximal current versus the graviton mass for various junction width $\ell$. }
\end{figure}

When the junction width is $\ell=2$, we show the relation between the tunneling current $J/T_c^2$ and the phase difference of the junction $\gamma$ in Fig.\ref{Jgamma}. By using the sinusoidal relation $J\approx J_{max} \sin(\gamma)$ to fit the data, we can figure out the maximal current $J_{max}$ for each $m$ and $\ell$. In Fig.\ref{Jmax}, we plot the relation between the maximal current $J_{max}$ to the graviton mass $m$ for various junction widths $\ell$. We find that for a fixed $\ell$, the maximal current will decrease as $m$ increases; Meanwhile for a fixed $m$, the maximal current decreases as well when $\ell$ increases. Physically, it means increasing the graviton mass (or equivalently increasing the momentum dissipation in the boundary) will suppress the tunneling between the two superconductors in the both sides of the junction.

\subsection{Coherence Length}
\begin{figure}[htbp]
\centering
\includegraphics[trim=0cm 0cm 0cm 0cm, clip=true,scale=0.27]{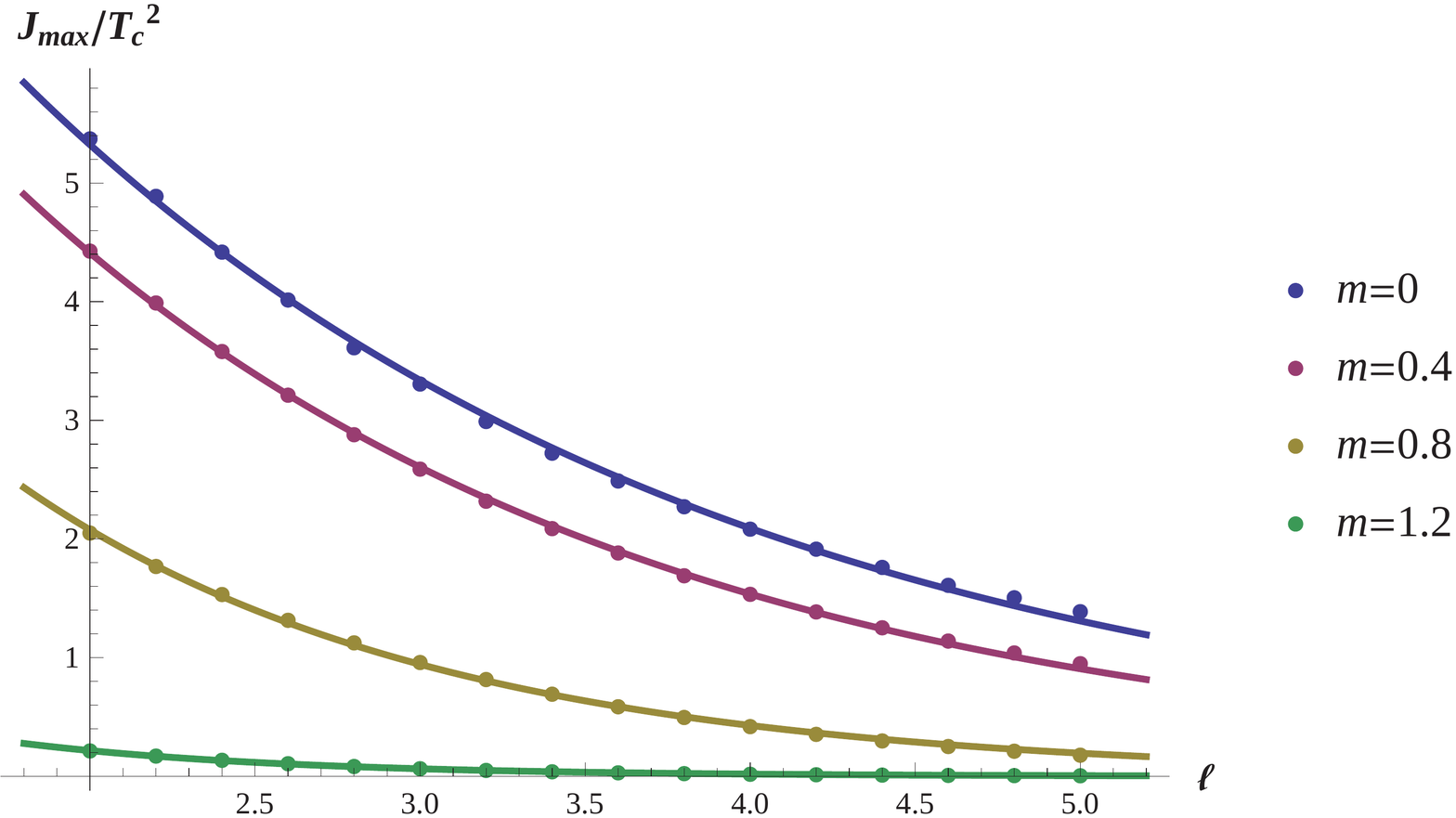}
\includegraphics[trim=0cm 0cm 0cm 0cm, clip=true,scale=0.23]{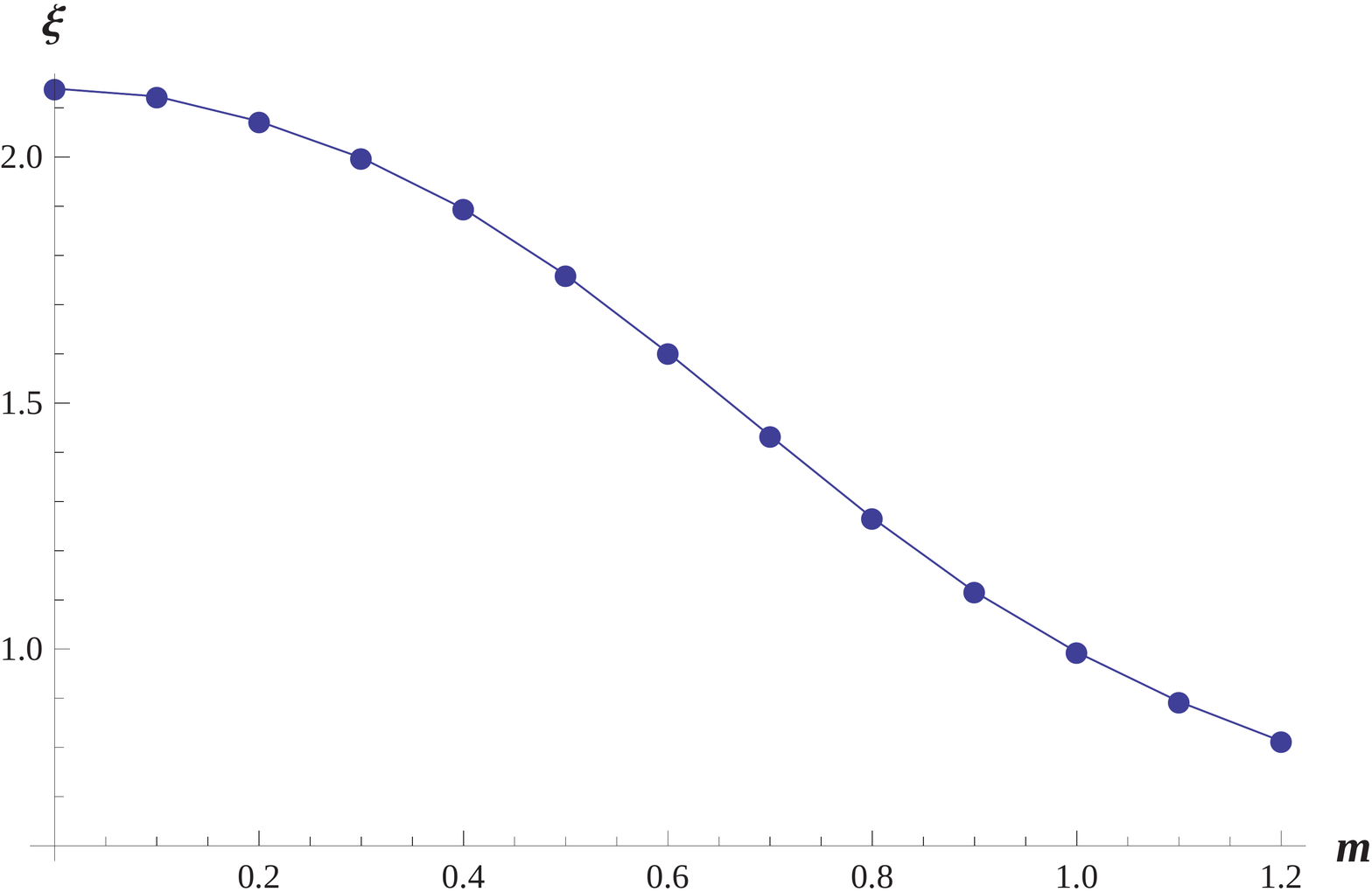}
\caption{\label{Ox0}(Left) Maximum current to the width of the junction $\ell$ for various $m$; (Right)Coherence length $\xi$ versus the graviton mass $m$ from the fitting \eqref{xi}. }
\end{figure}

From condensed matter physics \cite{tinkham}, there is a relation between the maximal current $J_{max}$ to the coherence length $\xi$ as
\be
\label{xi}
J_{max}/T_c^2&\approx& A e^{-\frac{\ell}{\xi}}
%\langle\O\rangle_{x=0}/T_c^2&\approx& A_2 e^{-\frac{\ell}{2\xi}}
\ee
This relation holds when $\ell\gg\xi$ where $\xi$ is the coherence length of the normal metal. We show the numerics and fittings of the relation \eqref{xi} in Fig.\ref{Ox0}. From the left side of Fig.\ref{Ox0}, we can see that for a fixed value of $m$, the maximal current has an exponential decay with respect to the width $\ell$. The fitted values of $\xi$ are shown on the right panel of Fig.\ref{Ox0}. We can see that the coherence length will decrease as $m$ increases, which means stronger momentum dissipation (breaking of translational symmetry) in the boundary field theory will reduce the coherence length $\xi$.

\section{Conclusions and Discussions}
\label{sect:con}

In this paper, we studied the SNS Josephson junction in the background of dRGT massive gravity. For a homogeneous chemical potential, we found that the greater the graviton mass was, the harder the normal metal-superconductor phase transition took place. From the aspects of holography we argued that the phase transition would be more difficult to happen if the momentum dissipation or breaking of translational symmetry was stronger in the boundary field theory. For the holographic SNS Josephson junction model, we obtained the familiar sinusoidal relation between the tunneling current and the phase difference across the junction. The maximal current would decrease by increasing the width of the junction which was similar to the previous studies. However, the more interesting thing was that by increasing the graviton mass in the bulk, the maximal current would decrease as well. Therefore, it indicated that stronger momentum dissipation would make the quantum tunneling in the Josephson junction harder to take place. By virtue of the relation between the maximal current and the coherence length, we found that the coherence length would decrease as well with respect to the graviton mass. Therefore, the momentum dissipation would also reduce the coherence length in the Josephson junction. We expect that this kind of relation between the maximal current, coherence length and the graviton mass (momentum dissipation in the boundary field theory) can be observed in the condensed matter experiments. It will also be interesting to find the analytic relation between the coherence length and the graviton mass.

\acknowledgments

  We are grateful to the KITPC's hospitality and partial support during completing this work. This work was partly supported by National Natural Science Foundation of China (NSFC) under grants No. 11105004, 11205097, 11575083, 11565017;  The Program for the Innovative Talents of Higher Learning Institutions of Shanxi; The Natural Science Foundation for Young Scientists of Shanxi Province, China (Grant No.2012021003-4); The Fundamental Research Funds for the Central Universities under grant No. NS2015073;  Shanghai Key Laboratory of Particle Physics and Cosmology under grant No. 11DZ2260700; The Open Project Program of State Key Laboratory of Theoretical Physics, Institute of Theoretical Physics, Chinese Academy of Sciences, China (No. Y5KF161CJ1);

\appendix
\section{Gauge-Invariance \& Constraint Equation}
\label{sect:appendix}
In this Appendix, we are going to show that for a generic Maxwell-complex scalar action \eqref{action2}, the constraint equation obtained from the imaginary part of the complex scalar field equation in fact can be derived from covariant derivative of the Maxwell equations. Therefore, the constraint equation actually is not independent from other equations.  The key points lie behind are the gauge-invariance of the EoMs. Assuming the gauge transformations (note that fields $|\psi|, \varphi$ and $M_\mu$ are real functions),
\be
\psi=|\psi|e^{i\varphi},\quad A_\mu=M_\mu+\partial_\mu\varphi.
\ee
The general complex scalar equation \eqref{eomA} becomes
\be\label{neweom1}
\left(\nabla_\mu-iM_\mu\right)\left(\nabla^\mu-iM^\mu\right)|\psi|-m_\psi^2|\psi|=0,
\ee
and the general Maxwell equation \eqref{eompsi} turns into
\be\label{neweom2}
\nabla_\nu F^{\nu\mu}=2M^\mu|\psi|^2.
\ee
Note that eqs.\eqref{neweom1} and \eqref{neweom2} are now equations of only gauge-invariant quantities, the phase $\varphi$ disappears from the EoMs. Furthermore, eq.\eqref{neweom1} can be decomposed into real and imaginary parts as
\be
\left(\nabla_\mu\nabla^\mu-M_\mu M^\mu-m_\psi^2\right)|\psi|-i\left(2M^\mu\nabla_\mu+(\nabla_\mu M^\mu)\right)|\psi|=0
\ee
The imaginary part of the complex scalar equation actually is the constraint equation (such as eq.\eqref{eom2}),
\be\label{cons}
2M^\mu\nabla_\mu|\psi|+(\nabla_\mu M^\mu)|\psi|=0
\ee
On the other hand, we can take covariant derivative of the Maxwell equation \eqref{neweom2}, and get (note that $ \nabla_\mu\left(\nabla_\nu F^{\nu\mu}\right)\equiv0$ since $F^{\nu\mu}$ is antisymmetric in the indices),
\be\label{newmax}
0\equiv \nabla_\mu\left(\nabla_\nu F^{\nu\mu}\right)=2\nabla_\mu\left(M^\mu|\psi|^2\right)\Rightarrow \left(\nabla_\mu M^\mu\right)|\psi|+2M^\mu\nabla_\mu|\psi|=0.
\ee
We can see that the right hand side of eq.\eqref{newmax} is exactly the constraint equation \eqref{cons}!

\end{CJK*}
\end{document}